\documentclass[a4paper,12pt]{article}
\usepackage[top=2cm, bottom=2cm, left=2cm, right=2cm]{geometry}

\newtheorem{thm}{Theorem}[section]

\newtheorem{rem}{Remark}[section]

\newtheorem{exmpl}{Example}[section]

\renewcommand{\arraystretch}{0.8}
\usepackage{mathtools}
\usepackage[numbers]{natbib}
\usepackage{graphicx}
\usepackage{amsmath}
\usepackage{amsfonts}
\usepackage{mathrsfs}
\usepackage{enumerate}
\usepackage{stfloats}
\usepackage{txfonts}
\usepackage{bm}
\usepackage{comment}
\usepackage{authblk}

\renewcommand{\arraystretch}{0.8}
\usepackage{arydshln}
\usepackage[ruled,vlined]{algorithm2e}

\begin{document}

 \title{A matrix pencil approach to the Morgan's problem}
\author{D. Vafiadis\thanks{d.vafiadis@econ.uoa.gr}}

\affil{
Department of Economics, Division of Mathematics-Informatics, National and Kapodistrian University of Athens,
 Athens,  Greece }

\date{}

\maketitle

\begin{abstract}
The problem of decoupling a nonsquare state space system by state feedback with singular input transformation  is considered.  The problem is solved by conducting a finite   search for decouplable square systems, appropriately derived from the original. Decoupling feedback on any of these systems defines the decoupling feedback for the original. The issue of fixed poles is also considered and the possibility of selecting the uncontrollable poles is investigated.

\end{abstract}

\section{Introduction}
 Morgan's problem is a long standing problem in control systems literature, which has attracted the interest of numerous researchers. In its general setting it is the problem of finding a state feedback law combined with an input transformation such that the resulting system has  diagonal and invertible transfer function. For the case where the original system is square, the problem is completely solved by following several approaches \cite{Falb-1967}, \cite{MorseWonham-1971}, \cite{Kous:DecI}. The general case is that of nonsquare systems  with  more inputs than outputs, where  the input transformation is necessarily singular in the sense that it corresponds to a nonsquare monic matrix. Most approaches on Morgan's problem \cite{Descusse-1984}, \cite{Descusse1988} \cite{HerreraH-Lafay1993}, \cite{KOUSSIOURIS1994}, \cite{Castaneda2012}  considered the derivation of  necessary and sufficient solvability conditions in terms of structural properties of the original system. The problem remained unsolved until 2017 where it was solved in the seminal paper  \cite{Kucera2017},  followed by  \cite{Kucera2019} and \cite{Kucera2021} where block decoupling and stability were also considered. The solvability conditions derived in the above paper involve the existence of certain lists of integers derived from system structural invariants, such as essential orders structure at infinity etc.

In the present paper an alternative approach to the solution of the decoupling problem is presented: The families of all square systems that could be  obtained from the original system by state feedback and singular input transformation, from now on referred to as {\it singular state feedback (SSF)},   are derived and, for each such family,  decouplability by state feedback and regular input transformation is tested. The classification of the families is based on the set of the controllability indices (CI). It is known -- see \cite{Loiseau1998}, \cite{Herrera2001} --  that the possible CI that can be obtained from a state space system by singular feedback can take specific values. This set of ``admissible'' CI is the starting point of the proposed method.
The candidate square systems for decoupling have CI from the admissible set. When the system is in Popov  canonical form, the controllability subspaces corresponding to these controllability  indices have basis matrix of a known structure \cite{Karc:BasisMatrix}. Using this basis matrix, the decouplability of all squared down systems, produced from the original can be easily  checked for all sets of admissible CI.
The decouplability check used  is the simple decoupling solvability condition of \cite{Kous:DecI}, which is based only on the CI and the output matrix of the system. Each set of CI characterises
the corresponding family of square systems obtained from the original system.
Going through all possible sets of CI we obtain all decouplable square systems generated from the original, if any exist.
The ``nice'' form of the basis matrix of controllability subspaces  enables the easy derivation of the initial singular feedback resulting in the square decouplable system. The decoupling pair of the latter, together with the preliminary feedback yield the solution of the problem. In the case where the  search
for decouplable systems returns nil, the problem is not solvable.
This search may return  multiple results i.e. systems of more than one set of CI. If the sum of the CI o the resulting system is less than the system dimension, input decoupling zeros are introduced in the final decoupled system. The assignment of these zeros is also considered in the present paper.

The paper is organised as follows: In Section \ref{ProblemStatement} the problem statement and   the necessary background  results on the decoupling of square systems   are given. Furthermore, an equivalent to the original, description of the system, convenient for the deployment of the rest of the paper is provided. In Section \ref{sectionSingularStateFeedback} it is shown that singular feedback augments the input--state pencil by a number of rows equal to the difference of the number  of inputs and outputs of the original system. The column minimal indices (CMI) of this pencil define the CI of the closed loop system, which in turn define the basis matrix of the corresponding CI of the closed loop system. From this basis matrix, a coordinate transformation on the state space is defined and can be used for checking the decouplability of the closed loop system. In Section
\ref{SolutionOfTheProblem} the procedure for searching for decouplable square systems obtained from the original is described in detail and, in case of existence of solution, the construction of SSF resulting in square system is described step by step. Finally, in Section \ref{FixedPoles} it is shown that  the fixed poles of the decoupled system are the uncontrollable poles introduced on the closed loop system due to singular feedback, and the already known fixed poles due to decoupling feedback \cite{WolovichFalb-1969}.

\section{Problem statement and background results}
\label{ProblemStatement}
Consider the controllable right invertible state -- space system
\renewcommand{\arraystretch}{0.7}
\setlength{\arraycolsep}{0.03in}%
\begin{equation}
\label{mp1} 
\dot{x}(t)=Ax(t)+Bu(t), \ y(t)=Cx(t)
\end{equation}
\noindent where $A \in \varmathbb{R}^{n \times n}$,  $B \in \varmathbb{R}^{n \times \ell}$,   $C \in \varmathbb{R}^{m \times n}$ and $m \leq \ell$.
Morgan's problem is stated as follows: Given system (\ref{mp1}),
find the  matrices $F \in \varmathbb{R}^{\ell \times n}$, $G \in \varmathbb{R}^{\ell \times m}$, $G$ 
monic  such that the system
\begin{equation}
\label{mp2} 
\dot{x}(t)=(A+BF)x(t)+BGu(t), \ y(t)=Cx(t)
\end{equation}
\medskip
\noindent
has diagonal transfer function $H_{F,G}(s)=diag(h_i(s)), i=1, \cdots,  \ell$. The above statement includes the case of square systems, equal number of inputs and outputs. This case was solved in \cite{Falb-1967}, \cite{Kous:DecI},  \cite{MorseWonham-1971}. In the present paper we are going to consider the nonsquare case where $m < \ell$. Feedback and input  transfornaton matrices $F$ and $G$  will be denoted   $(F,G)$.

The system matrix, \cite{Rosenbrock:StateSpaceMulti},   of (\ref{mp1}) is

\begin{equation}
\label{mp1_1}
T(s)=
\left[
\begin{array}{c;{2pt/2pt}c}
s I -A&-B\\
C&0
\end{array}
\right]
\end{equation}

\noindent
Let $\sigma_i,  i =1, \cdots ,\ell$ be the CI of system (\ref{mp1}) and $H(s)=N(s)D^{-1}(s)$ its transfer 
function, expressed in matrix fraction description (MFD) form. It is assumed that
the composite matrix 
$T(s)=[N^T(s),D^T(s)]^T$ is column reduced. Define the matrix $N_{\alpha} \in \varmathbb{R}^{m \times \ell}$ as follows
\begin{equation}
\label{mp3} 
N_{\alpha}=\left[N(s)\Delta(s) \right]_{hr}
\end{equation}
\medskip
\noindent
where $\Delta(s)=diag(s^{\sigma-\sigma_i})$, $ \sigma={\sf max}\{\sigma_i\}$ and  $\left[ \boldsymbol{\cdot} \right]_{hr}$ denotes the row highest order coefficient matrix, \cite{Kailath:book}.
\medskip
\noindent
Let
\begin{equation}
\label{eqsofs}
S(s)=diag \{ [1,s, \cdots, s^{\sigma_i-1}]^T \},  \ \ i =1, \cdots, \ell
\end{equation}
\medskip
\noindent
Then it is straightforward to see
(by using the Popov canonical form, \cite{Pop69:IrrMFD}, of a controllable system) that $N(s)=CS(s)$.
When the system is square ($m=\ell$), the solvability condition for the decoupling problem is given by the following result \cite{Kous:DecI}:

\begin{thm}
\label{thmDecouplingCondition}
The system  if system (\ref{mp1}) is square then it is decouplable by state feedback and regular input transformation if and only if the matrix $N_{\alpha} = [CS(s)\mbox{diag}\{s^{\sigma -  \sigma_i}\}]_{hr}$,
$\sigma = max\{\sigma_i\},  i =1, \cdots, \ell$ has full row rank. \hfill $\Box$
\end{thm}

Next we apply a series of transformations to the original system leading to an equivalent system in a form  convenient for the purposes of the paper.
We start from  (\ref{mp1_1}) and apply  similarity transformation $P$ leading to Popov
form, followed by input transformation $G_I$ such that the rows $\sigma_i$, $i=1, \cdots, \ell$ of the resulting input matrix are equal to $\underline{e}_i^T$,  the $i$-th component of the orthonormal basis of $\varmathbb{R}^{\ell}$. The resulting system is
\begin{equation}
\label{treq}
T_r(s)=
\left[
\begin{array}{c;{2pt/2pt}c}
s I -P^{-1}AP&-P^{-1}BG_I\\
CP&0
\end{array}
\right]=
\left[
\begin{array}{c;{2pt/2pt}c}
s I -A_r&-B_rG_I\\
C_r&0
\end{array}
\right]
\end{equation}

\noindent
Finally, we reorder the rows of the controllability pencil $[s I -A_r, \  -B_rG_I]$ of the above in an obvious way,   so that we have the system in form

\begin{equation}
\label{caf14}
T^d(s)=
\left[
\begin{array}{c;{2pt/2pt}c}
L(s)&0\\
sK -\Lambda & -I_\ell \\
C_r&0
\end{array}
\right]
\end{equation}
\noindent
where
\begin{equation}
\label{caf19_2}
L(s)=\mbox{diag}\{L_{\sigma_i}(s) \}
,
\sigma_1 \leq  \cdots \leq \sigma_\ell
,
L_{\sigma_i}(s)=
s\left[I_{\sigma_i-1}|0_{\sigma_i-1 \times 1} \right]-\left[
0_{\sigma_i-1\times 1}|I_{\sigma_i-1} \right]
\end{equation}
\begin{equation}
\label{caf19}
s K-  \Lambda=[s  K_1- \Lambda_1, \ldots,s  K_{\ell} -  \Lambda_{\ell}], \
K_i\in \mathbb R^{\ell \times \sigma_{i}}\,\ \Lambda_i \in \mathbb R^{\ell \times \sigma_{i}}
\end{equation}
\begin{equation}
\label{caf20}
K_i=
\left[
\begin{array}{cc}
0_{\ell \times (\sigma_{i}-1)},& \underline{e}_i
\end{array}
\right]
\end{equation}

\noindent
Note that $L(s)$ is actually the Kronecker canonical form of the {\it input--state pencil } $B^\perp(sI-A)$,  introduced in  \cite{Kar79:IEEPencil}, where $B^\perp$ is a left annihilator of $B$. Systems (\ref{mp1}) and (\ref{caf14}) have the same   decouplability properties, since (\ref{treq}) is related by similarity and regular input transformation to the original system. In the rest of the paper we are going to consider the Morgan's problem on
(\ref{caf14}) and we will return back to the original at the final stage of the problem solution.

\section{Singular state feedback and input -- state pencil}
\label{sectionSingularStateFeedback}
Sequential application of SSF  $(F_0,G_0)$ on (\ref{mp1}) and consequently on  (\ref{caf14}), leads to the following system matrix

\begin{equation}
\label{mpn1}
T^d_G(s)=
\left[
\begin{array}{c;{2pt/2pt}c}
L(s)&0\\
sK -\Lambda-F_0& -G_0
 \\ \hline
C_r&0
\end{array}
\right]
\end{equation}
\noindent
which can be further transformed by restricted system equivalence (RSE) transformations, \cite{Rosenbrock:StrProp}, as follows

\begin{equation}
\label{mpn2}
\left[
 \begin{array}{ccc}
I&0&0\\
0&\hat{P}&0\\
 0&0&I
 \end{array}
 \right]
\left[
\begin{array}{cc}
L(s)&0\\
sK -\Lambda& -G\\
C_r&0
\end{array}
\right]
\left[
\begin{array}{cc}
I&0\\
0&\hat{Q}
\end{array}
\right]=
\left[
\begin{array}{c|c}
L(s)&0\\
sK^a -\Lambda^a-F^a& 0\\ \hline
sK^b -\Lambda^b-F^b& -I_m\\
C_r&0
\end{array}
\right]=
\left[
\begin{array}{c|c}
 L^\alpha(s)&0\\
sK^b -\Lambda^b-F^b& -I_m\\ \hline
C_r&0
\end{array}
\right]
\end{equation}
where $F_0=[{F^a}^T,{F^b}^T]^T$. Pencil $L_\alpha(s)$ above is the input -- state pencil
 of the closed loop system and is obtained by $L(s)$ by row augmentation by the $\ell-m$ rows of the pencil $sK^a -\Lambda^a-F^a$.
Let 
\begin{equation}
 \label{aug0}
 L^\alpha(s)=
\left[
  \begin{array}{c}
  L(s) \\
 sK^a -\Lambda^a - F^a
  \end{array}
\right]
 \end{equation}

 \medskip
 \noindent
 Then
 \begin{equation}
 \label{aug0_1}
  \mathscr{N}_r(L_\alpha(s)) \subset \mathscr{N}_r(L(s))
 \end{equation}
 
 \medskip
 \noindent
where $\mathscr{N}_r(\boldsymbol{\cdot})$   denotes the right null space over the field of rational functons  $\varmathbb{R}(s)$. A polynomial basis matrix for $\mathscr{N}_r(L(s))$ is the matrix $S(s)$. If $\tilde \sigma_1, \cdots, \tilde \sigma_m$  are the CI
of the closed loop system (\ref{mp2}) then there exist constant nonsingular matrices $R$ and $Q$ of appropriate dimensions, with elemnts in $\varmathbb{R}$,
such that the system matrix (\ref{mpn2}) is transformed to

\begin{equation}
\label{nequ0_1}
\left[
 \begin{array}{ccc}
R&0&0\\
 0&I&0\\
 0&0&I
 \end{array}
 \right]
\left[
\begin{array}{cc}
 L^\alpha(s)&0\\
sK^b -\Lambda^b-F^b& -I_m\\
C_r&0
\end{array}
\right]
\left[
\begin{array}{cc}
Q&0\\
0&I
\end{array}
\right]
=
\left[
\begin{array}{cc}
 RL^\alpha(s)Q&0\\
(sK^b -\Lambda^b-F^b)Q& -I_m\\
C_rQ&0
\end{array}
\right]
\end{equation}
\noindent
with $RL_\alpha(s)Q$ in Kronecker canonical form i.e.
\begin{equation}
\label{nequ0}
RL_\alpha(s)Q
 =
 \left[
 \begin{array}{cc}
 D_f(s)&0\\
 0&\tilde L(s)
 \end{array}
\right]
\end{equation}
\medskip
\noindent
Matrix  $ \tilde L(s)$ above, is defined similarly to $L(s)$ in (\ref{caf19_2}) with respect to CI $\tilde \sigma_1, \cdots, \tilde \sigma_m$ and $D_f(s)$ is the block corresponding to the
finite elementary divisors (FED), see    \cite{Gantmache:Book},   of  $L_\alpha(s)$, i.e. the input decoupling zeros, \cite{Rosenbrock:StateSpaceMulti}, of the closed loop system.
Then a polynomial basis matrix for  $\mathscr{N}_r(RL_\alpha(s)Q)$ is
\begin{equation}
\label{nequ1}
S_f(s)=
 \left[
 \begin{array}{c}
 0\\
 \tilde S(s)
 \end{array}
\right] 
\end{equation}
\medskip
\noindent
i.e.
\begin{equation}
\label{nequ2}
 \left[
 \begin{array}{c}
  L(s)\\
  sK^a-\Lambda^a
 \end{array}
 \right]
 Q
 \left[
 \begin{array}{c}
 0\\
 \tilde S(s)
 \end{array}
\right]
=0
\end{equation}

\medskip
\noindent
Matrix $ \tilde S(s)$ is defined similarly to $S(s)$ defined in (\ref{eqsofs}) , with respect to the CI $\tilde \sigma_i$, $i=1,\cdots,m$ of the closed loop system.
Then, since a basis matrix of $\mathscr{N}_r(L(s))$ is $S(s)$, it follows from  (\ref{aug0_1})

\begin{equation}
\label{nequ3}
Q
 \left[
 \begin{array}{c}
 0\\
 \tilde S(s)
 \end{array}
\right]
=
S(s)V(s)
\end{equation}

\medskip
\noindent
where  $V(s) \in \varmathbb{R}^{\ell\times m}(s)$. Let

\begin{equation}
\label{nequ4}
Q=\left[Q_A, Q_B\right].
\end{equation}

\medskip
\noindent
Then (\ref{nequ3}) becomes
\begin{equation}
\label{nequ5}
Q_B \tilde S(s)=S(s)V(s)
\end{equation}
\medskip
\noindent
By equating the coefficients of like powers of $s$ above, we end up with $Q_B$ in the following form:

\begin{equation}
\label{nequ6}
\def\arraystretch{1.3}
Q_B=
\left[Q^1, \cdots, Q^m\right]=
\left[
 \begin{array}{c;{2pt/2pt}c;{2pt/2pt}c}
  Q_{1,1}&&Q_{1,m}\\
  \vdots&\cdots&\vdots\\
  Q_{\ell,1}&&Q_{m,m}\\
 \end{array}
\right]
\end{equation}

\medskip
\noindent
where

\begin{equation}
\label{nequ7}
\def\arraystretch{1.8}
Q_{i,j}=
 \left[
 \begin{array}{ccccccccccc}
 q^{ \sigma_i, \tilde \sigma_j}_1&& q^{ \sigma_i, \tilde \sigma_j}_2&\cdots&\cdots&\cdots& q^{ \sigma_i, \tilde \sigma_j}_{\sigma_j-\sigma_i}&0&\cdots&\cdots&0\\
 0&&q^{ \sigma_i, \tilde \sigma_j}_1& q^{ \sigma_i, \tilde \sigma_j}_2&&\cdots&\cdots& q^{ \sigma_i, \tilde \sigma_j}_{\sigma_j-\sigma_i}&0&\cdots&0\\
 \vdots&\ddots&\ddots&\ddots&&&\ddots&&&\ddots&\vdots\\
 0&\cdots&0&&q^{ \sigma_i, \tilde \sigma_j}_1& q^{ \sigma_i, \tilde \sigma_j}_2&\cdots&&\cdots&\cdots& q^{ \sigma_i, \tilde \sigma_j}_{\sigma_j-\sigma_i}
 \end{array}
 \right], \tilde \sigma_j \leq \sigma_i
\end{equation}

\medskip
\noindent
and
\begin{equation}
\label{nequ8}
Q_{i,j}={\Large{\mathbf{0}}}_{\sigma_i \times \tilde \sigma_j}, \tilde \sigma_j > \sigma_i
\end{equation}
\noindent
It can be easily verified that that
\noindent
\begin{equation}
\label{nequ8_1}
 \left[V(s)\right]_i=\left[ Q_B\tilde S(s)\right]_j,  j=1+ \sum_{k=1}^{i-1} \sigma_k, i=1,\cdots, \ell
\end{equation}

\noindent
 where $\left[ \boldsymbol{\cdot} \right]_i$ denotes the $i$--th row of a matrix.
From (\ref{nequ1}), (\ref{nequ0_1}), (\ref{nequ3}) and (\ref{nequ5}) it follows that a right MFD  of the transfer function of (\ref{nequ0_1}) with composite matrix $\tilde T(s)=[\tilde N^T(s),\tilde D^T(s)]^T$ is
\begin{equation}
\label{nequ8_2}
\tilde N(s) = CQ_B\tilde S(s), \ \tilde D(s) = (sK_b-\Lambda_b)Q_B\tilde S(s)=(sK_b-\Lambda_b)S(s) V(s)
\end{equation}
\noindent
The column block matrices $Q_j$ above, are basis matrices of CI of  dimension $\tilde \sigma_j$ of the pair $(A,B)$ (\cite{Karc:BasisMatrix} ) if and only if the conditions described in \cite{Warren-1975} and \cite{Karc:BasisMatrix} are satisfied, otherwise $Q^j$ is not of full rank and therefore Q is not of full rank. These conditions are given by the following theorem
\begin{thm}
\label{thmCSDimensions}
\cite{Warren-1975}, \cite{Karc:BasisMatrix}.
 An integer $\tilde \sigma_j$ can be the dimension of a CI of  $(A,B)$ if and only if
 \begin{equation}
 \label{nequ9}
  \sigma_k \leq \tilde \sigma_j \leq \sum_{i=1}^k \sigma_i
 \end{equation}
\medskip
\noindent
where $\sigma_k$ is the maximum minimal index of $(A,B)$ such that $\sigma_k \leq \tilde \sigma_j$.
 \hfill $\Box$
\end{thm}

\begin{rem}
\label{rem2}
It must me mentioned here that the term ``full rank'' refers to the generic rank, i.e.  that we may always select the parameters of matrices $Q_{i,j}$ in (\ref{nequ7})  such that  $Q_B$ is of full rank. \hfill $\Box$
\end{rem}

\noindent
The above theorem means that we can select $\tilde \sigma_j$ , $j=1, \cdots, m$ such that each one matrix  $Q^j$ has full column rank, but there is no guarantee that the whole matrix $Q_B$ has full column rank, that is, the controllability subspaces with basis matrices $Q^1, \cdots, Q^m$ defined in (\ref{nequ6}) are disjoint.  The condition on  $\tilde \sigma_j$ to result in $Q_B$ of full rank is given by the following Theorem
which is an alternative, equivalent, version of the result  presented in \cite{Herrera2001}:

\begin{thm}
\label{thmAdmissibleCI}
 Given the system with ordered CI $\sigma_i$, $i=1, \cdots, \ell$ the condition for an ordered set of integers $\{ \tilde \sigma_1, \cdots,  \tilde \sigma_m \}$ to result in $Q_B$ of full rank is
\begin{equation}
 \label{nequ10}
\sum_{j=1}^i {\tilde \sigma_j} \leq \sum_{j=1}^{k_i} \sigma_j
 \end{equation}
\medskip
\noindent
where $k_i={\sf max}\{j \ | \ \sigma_j \leq \tilde \sigma_i\}$, $i=1, \cdots, m$. \hfill $\Box$
\end{thm}

\section{The solution of the problem}
\label{SolutionOfTheProblem}
In this section the solution of the problem is obtained by systematically searching for solutions on ``input squared down'' systems obtained by singular input transformation. This search is finite and is described next.

Let $\mathscr{I}$ be the ordered list  of {\it admissible controllability indices}, i.e. the set of $m$-tuples of  CI that can be obtained by SSF from the original system. This set is fixed for $m$ fixed, and can be computed according to Theorem \ref{thmAdmissibleCI}. The $m$-tuples may contain repeated values of the CI, since the original system may have groups of equal CI. The cardinality of $\mathscr{I}$ is denoted by  $|\mathscr{I}|$ and the $i$--th element by  $\mathscr{I}(i) $.

The search for solution for the decoupling problem starts by picking a CI $m$-tuple  $\tilde \sigma_1, \cdots, \tilde  \sigma_m$  in $\mathscr{I}$.
For this set of CI the corresponding basis matrix $Q_B$  has the form   (\ref{nequ6}),  (\ref{nequ7})
and is of full column rank, since the CI of the candidate system are admissible.

The next step is to test the decouplability of the square system with the selected CI.
Decouplability of a square system depends only on the CI and output matrix C, as stated in  Theorem \ref{thmDecouplingCondition}.
If a system with the chosen set of CI is decouplable, then any similarity   equivalent system is also decouplable. Thus, instead of checking the decouplability of the original square system, we can check
the decouplability of the system with CI the chosen $m$-tuple, related to the original  with similarity tranformation Q, therefore with  output matrix $\hat{C}$ as in (\ref{nequ0_1}):
\begin{equation}
\label{nequ15}
\hat{C}=C_rQ=C\left[Q_A,Q_B \right]
\end{equation}
\noindent
In the above, there always exists  matrix $Q_A$  such that $Q=\left[Q_A,Q_B \right]$ is invertible, since $Q_B$ is monic. We construct matrix
\begin{equation}
 \label{nequ16}
 \hat N(s)=
\hat{C} \tilde S(s) diag(s^{\tilde \sigma-\tilde \sigma_i}), i=1,\cdots,m, \tilde \sigma=\mbox{max}\{\tilde \sigma_i \}
\end{equation}
\noindent
and check whether $Q_B$ can be selected such that the row high order coefficient matrix $N_{\alpha}$  (see Theorem \ref{thmDecouplingCondition})  of the above has full row rank.
This step can be performed starting from $N_\alpha$ obtained from (\ref{nequ16}) and searching  top to bottom for linearly dependent rows. Once such row is found  we eliminate the corresponding leading terms of the polynomial entries of $\hat N(s)$ by appropriate selection of the relevant parameters of $Q_B$, i.e.   equating the expressions forming the coefficients of the leading terms, to zero. Continuing this way with the new $N_\alpha$  and  $\hat N(s)$ we either get  $N_\alpha$ row reduced, or  $\hat N(s)$ becomes rank deficient which means that
the system is not decouplable with this set of CI.

If there is selection of  the parameters of $Q_B$  leading to $N_{\alpha}$ of full  rank, we are led to some constraints on  certain parameters of $Q_B$, i.e. fixed values on some parameters or algebraic constraints involving some of the parameters.
Next, we apply the constraints on $Q_B$  and check that
 $Q_B$ is of full column rank. An additional check that must be carried out, is that
\begin{equation}
  \label{nequ16_2}
  rank \left[(sK_b-\Lambda_b)Q_B\tilde S(s)\right]_{hc} =   rank \left[\tilde D(s)\right]_{hc}= m
 \end{equation}
\noindent
i.e. the denominator matrix $\tilde D(s)$  of the MFD of (\ref {nequ0_1}) is column reduced, so that Theorem \ref{thmDecouplingCondition} can be applied. If  $Q_B$ is rank deficient under these constraints we stop and pick another set of CI from $\mathscr{I}$.

 If the above three checks succeed, we can proceed and find the
appropriate  $F^a$ that results in $L^a(s)$ with the selected CMI corresponding to  the CI of the ``squared down system''. We may always assume that the blocks  $sK^a-\Lambda^a-F^a$ have the form  (\ref{caf19}), (\ref{caf20}) otherwise we can use coordinate transformations to obtain that form - see \cite{VafKarc95Nantes}. Note that in each row of  $sK^a-\Lambda^a-F^a$, entries with $s$ appear only once. The position of $s$ in each row corresponds to the last column of a CMI block $L_{\sigma_i}(s)$ of $L(s)$. If the row type  of   $sK^a-\Lambda^a-F^a$ is characterised   by the position of $s$ in it,  we may have $\binom{\ell}{\ell-m}$ possible $(\ell-m)$-tuples types of rows. Let the (lexicographically) ordered list  of these $(\ell-m)$-tuples be denoted by  $\mathscr{M}$ and its elements by $\mathscr{M}_j$,  $i=1, \cdots, \binom{\ell}{\ell-m}$.
  Now define  by   $M(s)=sK^a -\Lambda^a-F^a$ and denote by $\underline \mu^i(s)$ the rows of $M(s)$. Then $\underline \mu^i(s)$ have the form
\begin{equation}
 \label{nequ12}
\underline \mu^i(s)=\left[\mu_1^i,\mu_2^i, \cdots, \mu_{(\sum_{k=1}^j \sigma_j)-1}^i,  s+\mu^i_{\sum_{k=1}^j \sigma_j},\mu_{(\sum_{j=1}^i \sigma_i)+1}^i, \cdots, \mu^i_n \right], j = \mathscr{M}_j(i)
\end{equation}
\noindent
where $\mathscr{M}_j(i)$ is the $i$--th element of  $\mathscr{M}_j$.  Each $\mathscr{M}_i$ provides a different set of rows of $M(s)$. We start from $M(s)$ corresponding to $\mathscr{M}_1$. If, at a later stage of the procedure, this set of rows is found to be insufficient for providing decouplable system, we must  consider  $M(s)$ according to $\mathscr{M}_2$ and so on, until  either a solution is found, or all configurations for $M(s)$ are exhausted for the selected set of CI.
Then from  equations (\ref{nequ2}),(\ref{nequ3}) and (\ref{nequ4}) we have
\begin{equation}
  \label{nequ11}
\left[
  \begin{array}{c}
  L(s) \\
 sK^a -\Lambda^a-F^a
  \end{array}
\right]Q_B \tilde S(s)=0 \ \mbox{ or } \
\left[
  \begin{array}{c}
  L(s) \\
M(s)
  \end{array}
\right]Q_B \tilde S(s)=0
\end{equation}
\noindent
Our aim is to solve (\ref{nequ11}) with respect to  $\underline \mu^i_i(s)$. By construction, $Q_B$ is such that    $L(s) Q_B \tilde S(s)=0$. Then (\ref{nequ11})  reduces to
\begin{equation}
\label{nequ11_2}
M(s)\tilde S(s)=0
\end{equation}
\noindent
By equating the coefficients of the polynomial equations defined by  (\ref{nequ11}) to zero, we end up with the following systems of equations
\begin{equation}
\label{nequ13}
W_{\mu} \hat{\underline\mu}^i = Q^i_\mu
\end{equation}
\noindent
where $ \hat{\underline\mu}^i=\left[\mu_1^i, \cdots, \mu_n^i \right]$.
It can be easily verified that
\begin{equation}
\label{nequ14}
W_{\mu} = Q_B ^T
\end{equation}
\noindent
and since  $Q_B$ is monic by construction, (\ref{nequ13}) are always solvable. The solution of (\ref{nequ13}) depends on the parameters of $Q_B$. Thus, by assigning numeric values to these parameters while taking  care that the resulting $Q_B$ is monic, we obtain concrete values for the entries of $M(s)$. In some cases   it is necessary for some parameters of $Q_B$ to be equal to zero for (\ref{nequ13}) to be solvable. Then it is possible that $Q_B$ becomes column rank deficient. In that case  the selected $M(s)$ is not sufficient to provide the desired CI together with decouplability of the system, and we must change selection of $M(s)$ according to the next  $\mathscr{M}_i$.
The pair $\mathscr{I}_i$ and $\mathscr{M}_j$ will be referred to as {\it ``configuration'' } used for decouplability check.

Once a solution $M(s)$ is found, we may proceed to the next step. Matrix  of $sK^a -\Lambda^a$ consists  of the rows of $sK -\Lambda$ with index in $\mathscr{M}_i$, $i=1, \cdots, \ell-m$. We  apply a preliminary state feedback $F_0$ on the original system such that $sK^a -\Lambda^a-F^a$  is equal to the corresponding rows of $M(s)$ obtained in the previous steps. Thus, the components  $F^a$ and  $F^b$ of the preliminary feedback $F_0$ are selected as
\begin{equation}
\label{nequ18}
F^b=0, \ \
F^a = sK^a-\Lambda^a -M(s)
\end{equation}
\medskip
\noindent
Input transformation $G_0$ is constructed as follows: $G_0$ is an $\ell \times m$  matrix  obtained from the identity matrix $I_m$ by removing the columns with indices in $\mathscr{M}_i$, $i=1, \cdots, \ell-m$.

So far we have a system with  CI and output  martix satisfying the decouplability criteria, with system matrix
\begin{equation}
\label{tdfinal}
T_f^d(s)=
\left[
 \begin {array}{c|c}
L(s)&0\\
sK-\Lambda-F_0&-I_m \\ \hline
C&0
\end {array}
\right]
\end{equation}
\noindent
Reordering the rows of the above pencil such that the coefficient of $s$ is the identity matrix we get a state space system $(A_r,B_r,C_r)$ and applying   similarity transformation  Q to the latter, we obtain the following square system in state space form
\begin{equation}
\label{tfinal}
T_f(s)=
\left[
\begin {array}{c|c}
sI-A_f&-B_f\\ \hline
C_f&0
\end {array}
\right]
\end{equation}
\noindent
where
\begin{equation}
\label{finalmatrices}
\begin{array}{l}
B_f=Q^{-1}B_rG_iG_0=Q^{-1}P^{-1}BG_IG_0\\
\\
A_f=Q^{-1}(A_r-B_rF_0)Q=Q^{-1}P^{-1}(A-BF_0P^{-1})PQ\\
\\
C_f=C_rQ=CPQ
\end {array}
\end{equation}
\noindent
which is decouplable by state feedback and regular input transformation.
The pair $(A_f,B_f)$ is in controller canonical form, \cite{Kailath:book}.
What remains, is to find a feedback pair which decouples the system by using any of the well known decoupling techniques. Let the decoupling pair of (\ref{tfinal}) be $(F_f,G_f)$. Then the diagonal closed loop transfer function is
\begin{equation}
\label{diagtffinal}
H_{cl}(s)=
C_f(sI-A_f-B_fF_f)^{-1}B_fG_f
\end{equation}
\noindent
or, in terms of the original system matrices
\begin{equation}
\label{diagtforig}
H_{cl}(s)=
C\left( sI-A-B(G_IF_0P^{-1}+G_IG_0F_fQ^{-1}P^{-1}) \right)^{-1}\cdotp BG_iG_0G_f
\end{equation}
\noindent
Thus, the decoupling pair of the original system (\ref{mp1}) is
\begin{equation}
\label{finaldecpair}
F=G_I(F_0+G_0F_fQ^{-1})P^{-1}, \ G=G_IG_0G_f
\end{equation}

Below is the summary of the decoupling procedure in pseudocode format. The procedure terminates after a finite number of searches for decouplable systems among all configurations $\mathscr{M}_i$ for each one of the elements of $\mathscr{I}$. The maximum number of searches is $|\mathscr{I}|\cdot \binom{\ell}{\ell-m}$. It is mentioned here that we may get more than one solutions i.e. more than one closed loop systems differing on the sets of CI.

\bigskip
\noindent
{\bf Decoupling Procedure}
\par\vspace{-2.25mm}\noindent\rule{\textwidth}{0.5pt}

\bigskip
\noindent

{\bf Input:}System $(A,B,C)$

--Transform to  $(A_r,B_r,C_r)$ as in (\ref{treq})

--Reorder the equations to obtain system  (\ref{caf14})

--Calculate $\mathscr{I}$, $\mathscr{M}$

{\bf for} $i:= 1$ to $|\mathscr{I}|$ {\bf do}

\hspace*{0.15in}--Form  $Q_B$ corresponding to $\mathscr{I}(i)$  as in (\ref{nequ6}),(\ref{nequ7}),(\ref{nequ8}).

\hspace*{0.15in}--Test decouplability from $rank \left[\hat N(s)\right]_{hr}$ in  (\ref{nequ16}) and $rank \left[\tilde D(s)\right]_{hc}$  in (\ref{nequ16_2}).

\hspace*{0.15in} {\bf if} {System is decouplable and $Q_B$ has full rank}

\hspace*{0.25in}{\bf for}$j$ {\bf from} $1$ {\bf to} $\binom{\ell}{\ell-m}${\bf do}

            \hspace*{0.35in}--Form $M(s)$ with respect to $\mathscr{M}(j)$

            \hspace*{0.35in}--Select $Q_A$ such that $Q=[Q_A, Q_b]$ is invertible

            \hspace*{0.35in}--Form  (\ref{nequ13}) and check for constraints on parameters of $Q_B$

           \hspace*{0.35in} {\bf if} $Q_B$ has full rank under the above constraints and system is decouplable

                  \hspace*{0.50in}--Assign numeric values to $Q_B$

                  \hspace*{0.50in}--Solve (\ref{nequ13}) and form $M(s)$

                  \hspace*{0.50in}--Set $F_0$ and $G_0$ according to (\ref{nequ18})

                  \hspace*{0.50in}--Form system (\ref{tdfinal})

                  \hspace*{0.50in}--Reorder (\ref{tdfinal}) to get identity coefficient matrix of $s$ and

                    \hspace*{0.55in}apply similarity transformation $Q$ and get (\ref{tfinal})

                  \hspace*{0.50in}--Find decoupling pair $(F_f,G_f)$  for (\ref{tdfinal})

                  \hspace*{0.50in}--Set $(F,G)$ according to (\ref{finaldecpair}) and {\bf RETURN $(F,G)$ }

                  \hspace*{0.35in}{\bf end if}

                   \hspace*{0.25in}{\bf end for}

\hspace*{0.15in} {\bf end if}

{\bf end for}

{\bf RETURN NULL (NO SOLUTION EXISTS)}
\par\vspace{-2.25mm}\noindent\rule{\textwidth}{0.5pt}

\medskip

\noindent
The above can be stated formally as follows:
\begin{thm}
  Morgan's problem is solvable for system (\ref{mp1})  if and only if the above Decoupling Procedure returns at least one decoupling pair $(F,G)$.  \hfill $\Box$
\end{thm}

\noindent
Next, a fully worked example is given for illustrating the procedure.

\begin{exmpl} \cite{HerreraH-Lafay1993}.
 \label{example1}
 The given system is
\[
A= \left[ \begin {array}{ccccccccc} 0&0&0&0&0&0&0&0&0
\\ 0&0&0&0&0&0&0&0&0\\ 0&0&0&1&0&0
&0&0&0\\ 0&0&0&0&1&0&0&0&0\\ 0&0&0
&0&0&1&0&0&0\\ 0&0&0&0&0&0&0&0&0
\\ 0&0&1&0&0&0&0&1&0\\ 0&0&0&0&0&0
&0&0&1\\ 0&0&0&0&0&0&0&0&0\end {array} \right],
B=
\left[ \begin {array}{cccc} 1&0&0&0\\ 0&1&0&0
\\ 0&0&0&0\\ 0&0&0&0
\\ 0&0&0&0\\ 0&0&1&0
\\ 0&0&0&0\\ 0&0&0&0
\\ 0&0&0&1\end {array} \right]
C=
\left[ \begin {array}{ccccccccc} 1&0&0&0&0&0&0&0&0
\\ 0&1&0&0&0&0&0&0&0\\ 1&0&1&0&0&0
&0&0&0\end {array} \right]
\]
Apply  the similarity  and regular input transformations:
\begin{equation*}
\setlength{\arraycolsep}{0.01in}%
P=
\left[ \begin {array}{ccrrrrrrr}
1&0&0&0&0&0&0&0&0
\\ 0&1&0&0&0&0&0&0&0\\ 0&0&0&0&0&-
1&0&0&0\\ 0&0&0&0&0&0&-1&0&0\\ 0&0
&0&0&0&0&0&-1&0\\ 0&0&0&0&0&0&0&0&-1
\\ 0&0&-1&0&0&0&0&0&0\\ 0&0&0&-1&0
&1&0&0&0\\ 0&0&0&0&-1&0&1&0&0\end {array} \right],
G_I=
\left[
  \begin {array}{ccrr}
  1&0&0&0\\
   0&1&0&0\\
   0&0&0&-1\\
   0&0&-1&0
  \end {array}
\right]
\end{equation*}
The resulting system is in Popov canonical form:
\[A_r=
 \left[ \begin {array}{c;{2pt/2pt}c;{2pt/2pt}ccc;{2pt/2pt}cccc} 0&0&0&0&0&0&0&0&0
\\ \hdashline[2pt/2pt]
0&0&0&0&0&0&0&0&0\\ \hdashline[2pt/2pt]
0&0&0&1&0&0
&0&0&0\\
0&0&0&0&1&0&0&0&0\\
0&0&0
&0&0&0&0&1&0\\ \hdashline[2pt/2pt]
0&0&0&0&0&0&1&0&0
\\
0&0&0&0&0&0&0&1&0\\
0&0&0&0&0&0
&0&0&1\\
0&0&0&0&0&0&0&0&0\end {array} \right],
B_rG_I=P^{-1}BG_I=
\left[ \begin {array}{cccc}
  1&0&0&0\\ \hdashline[2pt/2pt]
  0&1&0&0\\\hdashline[2pt/2pt]
  0&0&0&0\\
  0&0&0&0\\
  0&0&1&0\\\hdashline[2pt/2pt]
  0&0&0&0\\
  0&0&0&0\\
  0&0&0&0\\
  0&0&0&1
\end {array} \right],
C_r=
 \left[ \begin {array}{c;{2pt/2pt}c;{2pt/2pt}ccc;{2pt/2pt}cccc} 1&0&0&0&0&0&0&0&0
\\ 0&1&0&0&0&0&0&0&0\\  1&0&0&0&0&-
1&0&0&0 \end {array} \right]
\]
\begin{equation*}
\left[
 \begin {array}{c}
L(s)\\
sK-\Lambda
\end {array}
\right]
=
 \left[
 \begin {array}{c;{2pt/2pt}c;{2pt/2pt}crr;{2pt/2pt}crrr}
0&0&s&-1&0&0&0&0&0\\
0&0&0&s&-1&0&0&0&0\\ \hdashline[2pt/2pt]
0&0&0&0&0&s&-1&0&0\\
0&0&0&0&0&0&s&-1&0\\
0&0&0&0&0&0&0&s&-1\\ \hline
s&0&0&0&0&0&0&0&0\\
0&s&0&0&0&0&0&0&0\\
0&0&0&0&s&0&0&-1&0\\
0&0&0&0&0&0&0&0&s
\end {array}
\right],
\end{equation*}
\noindent
The admissible CI set is $
 \mathscr{I}= \{(1, 1, 3), (1, 1, 4), (1, 1, 5), (1, 1, 6), (1, 1, 7), (1, 3, 4), (1, 3, 5),$  $ (1, 4, 4), (2, 3, 4)\}
$ and
$
\mathscr{M}=\{1,2,5,9\}  \ \mbox{i.e.} \ \mathscr{M}_1= (1), \mathscr{M}_2= (2), \mathscr{M}_3= (5), \mathscr{M}_4= (9)
$
corresponding to the following four possible configurations of the row of $M(s)$:
\[
\begin{array}{c}
\underline \mu^1(s)=[ \textbf{\textit{s}}+\mu_1^1,\mu_2^1,\mu_3^1,\mu_4^1,\mu_5^1,\mu_6^1,\mu_7^1,\mu_8^1,\mu_9^1]\\
\underline \mu^1(s)=[ \mu_1^1,\textbf{\textit{s}}+\mu_2^1,\mu_3^1,\mu_4^1,\mu_5^1,\mu_6^1,\mu_7^1,\mu_8^1,\mu_9^1]\\
\underline \mu^1(s)=[ \mu_1^1,\mu_2^1,\mu_3^1,\mu_4^1,\textbf{\textit{s}}+\mu_5^1,\mu_6^1,\mu_7^1,\mu_8^1,\mu_9^1]\\
\underline \mu^1(s)=[ \mu_1^1,\mu_2^1,\mu_3^1,\mu_4^1,\mu_5^1,\mu_6^1,\mu_7^1,\mu_8^1,\textbf{\textit{s}}+\mu_9^1]
\end{array}
\]
\noindent
We start with configuration
$\mathscr{I}(5)=(1, 1, 7)$ and $\mathscr{M}_1$ i.e. $M(s)=[ \textbf{\textit{s}}+\mu_1^1,\mu_2^1,\mu_3^1,\mu_4^1,\mu_5^1,\mu_6^1,\mu_7^1,\mu_8^1,\mu_9^1]$. Then
\begin{equation*}
Q_B=
\left[
 \resizebox{0.4 \textwidth}{!}{
 $
 \begin {array}{c;{2pt/2pt}c;{2pt/2pt}ccccccc}
q^{1,1}_{{1}}&q^{1,2}_{{1}}&
q^{1,3}_{{1}}&q^{1,3}_{{2}}&q^{1,3}_{{3}}&q^{1,3}_{{4}
}&q^{1,3}_{{5}}&q^{13}_{{6}}&q^{1,3}_{{7}}
\\ \hdashline[2pt/2pt]
q^{2,1}_{{1}}&q^{2,2}_{{1}}&q^{2,3}_{{1
,1}}&q^{2,3}_{{2}}&q^{2,3}_{{3}}&q^{2,3}_{{4}}&q^{2,3}_{
{5}}&q^{2,3}_{{6}}&q^{2,3}_{{7}}\\ \hdashline[2pt/2pt]
0&0&
q^{3,3}_{{1}}&q^{3,3}_{{2}}&q^{3,3}_{{3}}&q^{3,3}_{{4}}
&q^{3,3}_{{5}}&0&0\\
0&0&0&q^{3,3}_{{1}}&q^{3,3}_{{2}}&q^{3,3}_{{3}}&q^{3,3}_{{4}}&q^{3,3}_{{5}}
&0\\
0&0&0&0&q^{3,3}_{{1}}&q^{3,3}_{{2}}&
 q^{3,3}_{{3}}&q^{3,3}_{{4}}&q^{3,3}_{{5}}
\\ \hdashline[2pt/2pt]
0&0&q^{4,3}_{{1}}&q^{4,3}_{{2}}&q^{4,3}
_{{3}}&q^{4,3}_{{4}}&0&0&0\\
0&0&0&q^{4,3}_{
{1}}&q^{4,3}_{{2}}&q^{4,3}_{{3}}&q^{4,3}_{{4}}&0&0
\\
0&0&0&0&q^{4,3}_{{1}}&q^{4,3}_{{2}}&
q^{4,3}_{{3}}&q^{4,3}_{{4}}&0\\
0&0&0&0&0&
q^{4,3}_{{1}}&q^{4,3}_{{2}}&q^{4,3}_{{3}}&q^{4,3}_{{4}}
\end {array}
$
}
\right]
\end{equation*}
\noindent
Decouplability test: $\Delta(s)=diag \{s^6,s^6,1\}$
\[
\hat N(s)=
\left[
\resizebox{0.90 \textwidth}{!}{
$
\begin {array}{ccc} q^{1,1}_{{1}}{s}^{6}&q^{1,2}_{{1}}
{s}^{6}&q^{1,3}_{{7}}{s}^{6}+q^{1,3}_{{6}}{s}^{5}+q^{1,3}_{{
1,5}}{s}^{4}+q^{1,3}_{{4}}{s}^{3}+q^{1,3}_{{3}}{s}^{2}+
q^{1,3}_{{2}}s+q^{1,3}_{{1}}\\ q^{2,1}_{{1}}{
s}^{6}&q^{2,2}_{{1}}{s}^{6}&q^{2,3}_{{7}}{s}^{6}+q^{2,3}_{{1
,6}}{s}^{5}+q^{2,3}_{{5}}{s}^{4}+q^{2,3}_{{4}}{s}^{3}+q^{2,3
}_{{3}}{s}^{2}+q^{2,3}_{{2}}s+q^{2,3}_{{1}}
\\ q^{1,1}_{{1}}{s}^{6}&q^{1,2}_{{1}}{s}^{6}
&q^{1,3}_{{1}}-q^{4,3}_{{1}}+ \left( q^{1,3}_{{2}}-
q^{4,3}_{{2}} \right) s+ \left( q^{1,3}_{{3}}-q^{4,3}_{{3}}
 \right) {s}^{2}+ \left( q^{1,3}_{{4}}-q^{4,3}_{{4}} \right) {
s}^{3}+q^{1,3}_{{5}}{s}^{4}+q^{1,3}_{{6}}{s}^{5}+q^{1,3}_{{7}}{s}^{6}\end {array}
$
}
\right]
\]
\noindent
The above can have row highest order coefficient of full row rank only if
$q^{1,1}_{{1}}=q^{1,2}_{{1}}=
q^{1,3}_{{7}}=0$.
However in that case $Q_B$ is rank deficient since its two first columns are linearly dependent when $q^{1,1}_{{1}}=0$ and $q^{1,2}_{{1}}=0$ .

\noindent
Following the same procedure, we find that all elements of $\mathscr{I}$ except $(1,4,4)$ lead to non decouplable square systems. For the configuration  $\mathscr{I}(8)=(1,4,4)$ and $\mathscr{M}_1$ we have
\begin{equation*}
Q_B=
\left[ \begin {array}{c;{2pt/2pt}cccc;{2pt/2pt}cccc} q^{1,1}_{{1}}&q^{1,2}_{{1}}&
q^{1,2}_{{2}}&q^{1,2}_{{3}}&q^{1,2}_{{4}}&q^{1,3}_{{1}
}&q^{1,3}_{{2}}&q^{1,3}_{{3}}&q^{1,3}_{{4}}\\ \hdashline[2pt/2pt]
q^{2,1}_{{1}}&q^{2,2}_{{1}}&q^{2,2}_{{2}}&q^{2,2}_{{3}}&q^{2,2}_{{4}}&q^{2,3}_{{1}}&q^{2,3}_{{2}}&q^{2,3}_{{3}}&q^{2,3}_{{4}}\\ \hdashline[2pt/2pt]
0&q^{3,2}_{{1}}&q^{3,2}_{{2}}&0&0&q^{3,3}_{{1}}&q^{3,3}_{{2}}
&0&0\\
0&0&q^{3,2}_{{1}}&q^{3,2}_{{2}}&0&0&q^{3,3}_{{1}}&q^{3,3}_{{2}}&0\\
0&0&0&q^{3,2}_{{1}}&q^{3,2}_{{2}}&0&0&q^{3,3}_{{1}}&q^{3,3}_{{2}}\\ \hdashline[2pt/2pt]
0&q^{4,2}_{{1}}&0&0&0&q^{4,3}_{{1}}&0&0&0\\
0&0&q^{4,2}_{{1}}&0&0&0&q^{4,3}_{{1}}&0&0\\
0&0&0&q^{4,2}_{{1}}&0&0&0&q^{4,3}_{{1}}&0\\
0&0&0&0&q^{4,2}_{{1}}&0&0&0&q^{4,3}_{{1}}
\end {array} \right]
\end{equation*}
\begin{equation*}
\hat{C}=C Q_B=
 \left[ \begin {array}{ccccccccc} q^{1,1}_{{1}}&q^{1,2}_{{1}}&
q^{1,2}_{{2}}&q^{1,2}_{{3}}&q^{1,2}_{{4}}&q^{1,3}_{{1}
}&q^{1,3}_{{2}}&q^{1,3}_{{3}}&q^{1,3}_{{4}}
\\
q^{2,1}_{{1}}&q^{2,2}_{{1}}&q^{2,2}_{{2}}&q^{2,2}_{{3}}&q^{2,2}_{{4}}&q^{2,3}_{{1}}&q^{2,3}_{
{1,2}}&q^{2,3}_{{3}}&q^{2,3}_{{4}}\\
q^{1,1}_{{1}}&q^{1,2}_{{1}}-q^{4,2}_{{1}}&q^{1,2}_{{2}}&q^{1,2}_{{3}}&q^{1,2}_{{4}}&q^{1,3}_{{1}}-q^{4,3}_{{1}}
&q^{1,3}_{{2}}&q^{1,3}_{{3}}&q^{1,3}_{{4}}\end {array}
 \right]
\end{equation*}
\noindent
Note that here $Q=Q_B$. For testing decouplability we have $\Delta(s)=diag \{s^3,1,1\}$ and
\begin{equation*}
\hat{N}(s)=
 \left[ \begin {array}{ccc} q^{1,1}_{{1}}{s}^{3}&{s}^{3}q^{1,2}_
{{1,4}}+{s}^{2}q^{1,2}_{{3}}+sq^{1,2}_{{2}}+q^{1,2}_{{1}}&
{s}^{3}q^{1,3}_{{4}}+{s}^{2}q^{1,3}_{{3}}+sq^{1,3}_{{2}}+{
\it q13}_{{1}}\\
q^{2,1}_{{1}}{s}^{3}&{s}^{3}
q^{2,
2}_{{4}}+{s}^{2}q^{2,2}_{{3}}+sq^{2,2}_{{2}}+q^{2,2}
_{{1}}&{s}^{3}q^{2,3}_{{4}}+{s}^{2}q^{2,3}_{{3}}+sq^{2,3}_
{{2}}+q^{2,3}_{{1}}\\
q^{1,1}_{{1}}{s}^{3}
&{s}^{3}q^{1,2}_{{4}}+{s}^{2}q^{1,2}_{{3}}+sq^{1,2}_{{2}}+
q^{1,2}_{{1}}-q^{4,2}_{{1}}&{s}^{3}q^{1,3}_{{4}}+{s}^{2}q^{13}_{{3}}+sq^{1,3}_{{2}}+q^{1,3}_{{1}}-q^{4,3}_{{1}
}\end {array} \right]
\end{equation*}

\noindent
\begin{equation*}
N_{\alpha}=
  \left[ \begin {array}{ccc} q^{1,1}_{{1}}&q^{1,2}_{{4}}&
q^{1,3}_{{4}}\\  q^{2,1}_{{1}}&q^{2,2}_{{4}}&q^{2,3}_{{4}}\\  q^{1,1}_{{1}}&q^{1,2}_{{4
}}&q^{1,3}_{{4}}\end {array} \right]
\end{equation*}
\noindent
Matrix $N_{\alpha}$ is rank deficient. Now set $
 q^{1,1}_{1}=q^{1,2}_{4}=q^{1,3}_{4}= q^{1,2}_{3}=q^{1,3}_{3}=q^{1,2}_{2}=q^{1,3}_{2}=0
$ on $Q_B$, and
\begin{equation*}
N_{\alpha}=
\left[ \begin {array}{ccc} 0&q^{1,2}_{{1}}&q^{1,3}_{{1}}
\\  q^{2,1}_{{1}}&q^{2,2}_{{4}}&q^{2,3}_{{1
,4}}\\  0&q^{1,2}_{{1}}-q^{4,2}_{{1}}&
q^{1,3}_{{1}}-q^{4,3}_{{1}}\end {array} \right]
\end{equation*}
\medskip
\noindent
has rank 3 for generic values of the parameters. Matrix
\begin{equation*}
\left[\tilde D(s) \right]_{hc}=
\left[ \begin {array}{ccc}
q^{2,1}_{{1}}&q^{2,2}_{{4}}&
q^{2,3}_{{4}}\\
0&q^{3,2}_{{2}}&q^{3,3}_{{2}}\\
0&q^{4,2}_{{1}}&q^{4,3}_{{1}}
\end {array} \right]
\end{equation*}
\noindent
is also of full rank. Equations (\ref{nequ11_2}) are
\begin{equation*}
 \begin {array}{c}
  \mu^1_2\,q^{2,1}_{{1}}=0\\
  q^{1,2}_{{1}}\mu^1_1+\mu^1_2\,q^{2,2}_{{1}}+\mu^1_3\,q^{3,2}_{{1}}+
\mu^1_6\,q^{4,2}_{{1}}=0\\
 \mu^1_2\,q^{2,2}_{{2}}+\mu^1_3\,q^{3,2}_{{2}}+\mu^1_4\,
q^{3,2}_{{1}}+\mu^1_7\,q^{4,2}_{{1}}+q^{1,2}_{{1}}=0\\
 \mu^1_2\,q^{2,2}_{{3}}+\mu^1_4\,
q^{3,2}_{{2}}+\mu^1_5\,q^{3,2}_{{1}}+\mu^1_8\,q^{4,2}_{{1}} \mu^1_8=0\\
 \mu^1_2\,q^{2,2}_{{4}}+\mu^1_5\,q^{3,2}_{{2}}+\mu^1_9\,q^{4,2
}_{{1}} \mu^1_9 =0\\
q^{1,3}_{{1}}\mu^1_1+\mu^1_
2\,q^{2,3}_{{1}}+\mu^1_3\,q^{3,3}_{{1}}+\mu^1_6\,q^{4,3}_{{1}}=0\\
q^{2,3}_{{2}}+\mu^1_3\,q^{3,3}_{{2}}+\mu^1_4\,q^{3,3}_{{1}}+\mu^1_7\,
 q^{4,3}_{1}+q^{1,3}_{{1}}  =0\\
 \mu^1_2\,q^{2,3}_{{3}}+\mu^1_4\,q^{3,3}_{{2}}+\mu^1_5\,q^{3,3
}_{{1}}+\mu^1_8\,q^{4,3}_{{1}} \mu^1_8=0\\
 \mu^1_2\,q^{2,3}_{{4}}+\mu^1_5\,q^{3,3}_{{2}}+\mu^1_9\,q^{4,3}_{{1}} \mu^1_9=0
 \end {array}
\end{equation*}
\noindent
or, in matrix form $W_{\mu} \hat{\underline\mu}^1 =Q_B^T \hat{\underline\mu}^1= Q^1_\mu$ i.e.
\begin{equation*}
\left[ \begin {array}{ccccccccc}
0&q^{2,1}_{{1}}&0&0&0&0&0&0&0\\
q^{1,2}_{{1}}&q^{2,2}_{{1}}&q^{3,2}_{{1}}&0&0&q^{4,2}_{{1}}&0&0&0\\
0&q^{2,2}_{{2}}&q^{3,2}_{{2}}&q^{3,2}_{{1}}&0&0&q^{4,2}_{{1}}&0&0\\
0&q^{2,2}_{{3}}&0&q^{3,2}_{{2}}&q^{3,2}
_{{1}}&0&0&q^{4,2}_{{1}}&0\\ 0&q^{2,2}_{{4
}}&0&0&q^{3,2}_{{2}}&0&0&0&q^{4,2}_{{1}}\\
q^{1,3}_{{1}}&q^{2,3}_{{1}}&q^{3,3}_{{1}}&0&0&q^{4,3}_{{1
,1}}&0&0&0\\ 0&q^{2,3}_{{2}}&q^{3,3}_{{2}}&q^{3,3}_{{1}}&0&0&q^{4,3}_{{1}}&0&0\\ 0&
q^{2,3}_{{3}}&0&q^{3,3}_{{2}}&q^{3,3}_{{1}}&0&0&q^{4,3}_{{1
}}&0\\ 0&q^{2,3}_{{4}}&0&0&q^{3,3}_{{2}}&0&0
&0&q^{4,3}_{{1}}\end {array} \right]
\left[ \begin {array}{c}
\mu^1_1\\
\mu^1_2\\
\mu^1_3\\
\mu^1_4\\
\mu^1_5\\ \mu^1_6
\\ \mu^1_7\\ \mu^1_8\\ \mu^1_9
\end {array} \right]
=
 \left[ \begin {array}{c} 0\\ 0\\
-q^{1,2}_{{1}}\\ 0\\ 0
\\ 0\\ -q^{1,3}_{{1}}
\\ 0\\ 0\end {array} \right]
\end{equation*}

\noindent
Setting
$
 q^{3,2}_{2}=q^{3,3}_{2}=q^{4,2}_{1}= q^{3,2}_{1}=q^{3,3}_{1}=q^{1,2}_{1}=q^{1,3}_{1}=1
,
q^{4,3}_{1}=2, q^{3,2}_{2}=q^{3,3}_{2}=q^{4,2}_{1}= q^{3,2}_{1}=q^{3,3}_{1}=q^{1,2}_{1}=q^{1,3}_{1}=1
,
q^{2,1}_{1}=1, q^{2,2}_{1}=q^{2,2}_{2}=q^{2,2}_{3}= q^{2,2}_{4}=q^{2,3}_{1}=q^{2,3}_{2}=q^{2,3}_{3}=q^{2,3}_{4}=0
$,
the above equation has solution

\begin{equation*}
\left[
\mu^1_1,
\mu^1_2,
\mu^1_3,
\mu^1_4,
\mu^1_5,
\mu^1_6,
\mu^1_7,
\mu^1_8,
\mu^1_9
\right]
=
\left[
1,
0,
-1,
0,
0,
0,
 0, 0, 0
\right]
\end{equation*}
\noindent
and
\begin{equation*}
Q_B=
\left[ \begin {array}{c;{2pt/2pt}cccc;{2pt/2pt}cccc} 0&1&0&0&0&1&0&0&0\\ \hdashline[2pt/2pt]
1&0&0&0&0&0&0&0&0\\ \hdashline[2pt/2pt]
0&1&1&0&0&1&1&0&0\\
0&0&1&1&0&0&1&1&0\\
0&0&0&1&1&0&0&1&1\\ \hdashline[2pt/2pt]
0&1&0&0&0&2&0&0&0\\
0&0&1&0&0&0&2&0&0\\
0&0&0&1&0&0&0&2&0\\
0&0&0&0&1&0&0&0&2
\end {array} \right]
\end{equation*}
\noindent
Furthermore rank of matrices in $N_{\alpha}$ and $\left[\tilde D(s) \right]_{hc}$ is 3. For configuration $\mathscr{I}(8)$ and $\mathscr{M}_1$ , we have  $M(s)=[s+1,0,-1,0,0,0,0,0,0]$, $sK^a-\Lambda^a=[s,0,0,0,0,0,0,0,0]$ and the preliminary feedback pair $F_0$, $G_0$ obtained from (\ref{nequ18}) is
\begin{equation*}
F_0=
\left[ \begin {array}{ccccccccc}
-1&0&1&0&0&0&0&0&0\\ \hdashline[2pt/2pt]
0&0&0&0&0&0&0&0&0\\
0&0&0&0&0&0&0&0&0\\
0&0&0&0&0&0&0&0&0
\end {array} \right], \
G_0=
\left[ \begin {array}{ccc}
 0&0&0\\
 1&0&0\\
 0&1&0\\
 0&0&1
 \end {array}
\right]
\end{equation*}
\noindent
The preliminary square system in the form (\ref{mpn2}) is
\begin{equation*}
\left[
 \begin {array}{cc}
L(s)&0\\
sK-\Lambda-F_0&-I_m
\end {array}
\right]
=
 \left[
 \begin {array}{c;{2pt/2pt}c;{2pt/2pt}crr;{2pt/2pt}crrr|rrr}
0&0&s&-1&0&0&0&0&0&0&0&0\\
0&0&0&s&-1&0&0&0&0&0&0&0\\ \hdashline[2pt/2pt]
0&0&0&0&0&s&-1&0&0&0&0&0\\
0&0&0&0&0&0&s&-1&0&0&0&0\\
0&0&0&0&0&0&0&s&-1&0&0&0\\
s+1&0&-1&0&0&0&0&0&0&0&0&0\\\hline
0&s&0&0&0&0&0&0&0&-1&0&0\\
0&0&0&0&s&0&0&-1&0&0&-1&0\\
0&0&0&0&0&0&0&0&s&0&0&-1
\end {array}
\right],
\end{equation*}
\noindent
Now, reordering the rows of the above pencil such that the coefficient of $s$ is the identity matrix we get the system
\[
\left[
\begin {array}{c|c}
sI-A_r-BrF_0&B_rG_0\\ \hline
C_r&0
\end {array}
\right]
\]
\noindent
Similarity transformation Q on the above results to the system  with matrix
\[
\left[
\begin {array}{c|c}
sI-A_f&B_f\\ \hline
C_f&0
\end {array}
\right]
=
\left[
\begin {array}{c;{2pt/2pt}rrrc;{2pt/2pt}crrc|crr} s&0&0&0&0&0&0&0&0&1&0&0
\\ \hdashline[2pt/2pt]
0&s&-1&0&0&0&0&0&0&0&0&0\\
0&0&
s&-1&0&0&0&0&0&0&0&0\\
0&0&0&s&-1&0&0&0&0&0&0&0
\\
0&0&0&-2&2+s&0&0&-4&2&0&2&-1\\ \hdashline[2pt/2pt]
0&0&0&0&0&s&-1&0&0&0&0&0\\ 0&0&0&0&0&0&s&-1&0&0&0&0
\\ 0&0&0&0&0&0&0&s&-1&0&0&0\\ 0&0&0
&1&-1&0&0&2&-1+s&0&-1&1\\ \hline
0&1&0&0&0&1&0&0&0&0&0&0
\\
1&0&0&0&0&0&0&0&0&0&0&0\\
0&0&0
&0&0&-1&0&0&0&0&0&0
\end {array}
\right]
\]
\noindent
Notice that $(A_f,B_f)$ is in controller form revealing the CI. The above system is decouplable. The feedback pair
\[
F_f=
\left[ \begin {array}{rcrrccrrc} -3&0&0&0&0&0&0&0&0
\\ \noalign{\medskip}0&3&-2&-1&1&3&-2&-2&1\\ \noalign{\medskip}0&3&-2&0
&0&4&-3&0&0\end {array} \right],
G_f=
\left[ \begin {array}{ccr} 0&1&0\\ \noalign{\medskip}1&0&0
\\ \noalign{\medskip}1&0&-1\end {array} \right]
\]
\noindent
yields  the diagonal transfer matrix

\[
H_{cl}(s)= \mbox{diag} \Biggl\{  \left( {s}^{4}+2\,s-3 \right)^{-1}, \left( s+3 \right)^{-1}, \left( {s}^{4}+s-1 \right)^{-1} \Biggr\}
\]
\noindent
Then, the  decoupling pair of the original system  (see (\ref{finaldecpair})) is
\[
F=
\left[ \begin {array}{rrrrrcrrc} -1&0&0&0&0&0&-1&0&0
\\  0&-3&0&0&0&0&0&0&0\\  -3&0&1&-1&0
&0&-1&0&0\\  -4&0&-1&1&-1&0&-1&-1&1\end {array}
 \right],
 G=\left[ \begin {array}{rcc} 0&0&0\\  0&1&0
\\  -1&0&1\\  -1&0&0\end {array}
 \right]
\]
\end{exmpl}

\section{Fixed poles}
\label{FixedPoles}

In Section \ref{sectionSingularStateFeedback}  is was mentioned (see (\ref{nequ0}))  that the FED of $L_a(s))$ are the input decoupling zeros of the closed loop  system, if a solution exists. The decoupling zeros cannot be shifted by state feedback and belong to the set of fixed pols of the decoupled system. The rest of the fixed poles are those poles correspond to the constraints of decoupling. It is well known that the fixed poles of a decoupled system are unobservable i.e. they  are output decoupling zeros. In this section the problem of selecting input decoupling zeros, when it is possible is discussed and a characterisation of the output decoupling zeros is given.

Since of the original system is considered to be controllable, from  (\ref{nequ0}) and (\ref{nequ4}) we see that the closed loop system may have input decoupling zeros only if the sum of the CI $\tilde \sigma_i$ is less than the dimension $n$ i.e.
\begin{equation}
 \label{eqfp1}
 \sum_{i=1}^m \tilde \sigma_i < n
 \end{equation}
\noindent
Then the dimension of the FED block $D_f(s)$ in (\ref{nequ0}) is $n-\sum_{i=1}^m \tilde \sigma_i$. Consider (\ref{nequ0}) and write $R^{-1}$ as
\begin{equation}
 \label{eqfp2}
R^{-1}=\left[R_A,R_B\right]
 \end{equation}
\noindent
Then
\begin{equation}
 \label{eqfp3}
L_a(s)\left[Q_A,Q_B\right]=\left[R_AD_f(s),R_B\tilde L(s)\right]
\end{equation}
\noindent
Now let $R^\perp$ and $R^\dag$   be a left annihilator and left inverse of  $R_B$, respectively. Then
\begin{equation}
 \label{eqfp4}
 \left[
 \begin{array}{c}
  R^\perp \\
  R^\dag
\end{array}
\right]
 L_a(s)\left[Q_A,Q_B\right]=
\left[
\begin{array}{cc}
  R^\perp R_AD_f(s) & 0\\
  R^\dag R_AD_f(s)& \tilde L(s)
\end{array}
\right]
\end{equation}
\noindent
The above means that the FED of $L_a(s)$ are those of
$R^\perp R_AD_f(s)$. Note that $R^\perp R_A$ is square and invertible.
When (\ref{eqfp1}) holds,  the number of equations in (\ref{nequ13}) is less than the number of unknowns
which can be used in order to select the input decoupling zeros (the zeros of $R^\perp R_AD_f(s)$) .
The fixed poles due to decoupling are the zeros of the polynomial (see
\cite{WolovichFalb-1969})
\begin{equation}
  \label{eqfp5}
\frac{\left|C_f \cdot S_f(s) \right|}{\prod_{i=1}^m\xi_i(s)}
\end{equation}
\noindent
where $S_f(s)$ is defined in (\ref{nequ1}) and  $\xi_i(s)$, $i=1, \cdots, m$ is the
  the greatest common divisors of the elemants  of the $i$--th row of $C_f \cdot S_f(s)$. Both types of fixed poles, the input decoupling zeros and the roots of (\ref{eqfp5}) depend on the CI of the closed loop system. Thus, in the case where decoupling can be achieved by different sets of CI, the resulting fixed poles are different. Finally, in the case where the sum of closed loop CI is equal to $n$, the resulting system is controllable and no uncontrollable fixed poles exist.
\begin{exmpl}
 \label{example2}
 Consider the system with system matrix
\begin{equation*}
\label{exmpl2_1}
T(s)=
\left[
\begin{array}{c|c}
s I -A&-B\\
C&0
\end{array}
\right]=
 \left[ \begin {array}{c;{2pt/2pt}cr;{2pt/2pt}cr;{2pt/2pt}rr;{2pt/2pt}cc|rrrrr}
 s&0&0&0&0&-1&0&0&0&-1&0&0&0&0\\  \hdashline[2pt/2pt]
0&s&-1&0&0&0&0&0&0&0&0&0&0&0\\
0&1&s&0&0&0&0&0&0&0&-1&0&0&0\\  \hdashline[2pt/2pt]
0&0&0&s&-1&0&0&0&0&0&0&0&0&0\\
0&0&-1&0&s&0&1&0&0&0&0&-1&0&0\\ \hdashline[2pt/2pt]
0&0&0&0&0&s&-1&0&0&0&0&0&0&0\\
0&0&0&0&0&0&s&0&0&0&0&0&-1&0\\
0&0&0&0&0&0&0&s&-1&0&0&0&0&0\\
0&0&0&0&0&0&0&0&s+1&0&0&0&0&-1\\ \hline
0&1&0&0&0&0&1&1&1&0&0&0&0&0\\
1&0&1&1&0&0&0&0&1&0&0&0&0&0\\
1&0&0&0&0&0&0&0&1&0&0&0&0&0
\end {array} \right]
\end{equation*}
\noindent
For this system $\sigma_1=1$, $\sigma_2=\sigma_3=\sigma_4=\sigma_5=2$,
 $
 \mathscr{I}= \{(1, 2, 2)$, $(1, 2, 3)$, $(1, 2, 4)$, $ (1, 2, 5), (1, 2, 6), $ $(1, 3, 3),$ $ (1, 3, 4)$, $(1, 3, 5)$, $(1, 4, 4), $ $(2, 2, 2), (2, 2, 3)$, $(2, 2, 4) $, $(2, 2, 5)$, $(2, 3, 3)$, $(2, 3, 4)$, $(3, 3, 3)\}$ and
$\mathscr{M}=\{(1, 3),$ $ (1, 5)$, $ (1, 7)$, $(1, 9)$, $(3, 5), (3, 7), $ $(3, 9), (5, 7), $ $(5, 9), (7, 9)\}$. For the configuration $\mathscr{I}(11)=(2,2,3)$, $\mathscr{M}_2=(1,5)$ we have
\[
\begin{array}{c}
\underline \mu^1(s)=[ \textbf{\textit{s}}+\mu_1^1,\mu_2^1,\mu_3^1,\mu_4^1,\mu_5^1,\mu_6^1,\mu_7^1,\mu_8^1,\mu_9^1]\\
\underline \mu^2(s)=[ \mu_1^2,\mu_2^2,\mu_3^2,\mu_4^2,\textbf{\textit{s}}+\mu_5^2,\mu_6^2,\mu_7^2,\mu_8^2,\mu_9^2]\\
\end{array}
\]
\[
Q_B=
 \left[ \begin {array}{cc;{2pt/2pt}cc;{2pt/2pt}ccc} q^{1,1}_{{1}}&q^{1,1}_{{2}}&
q^{1,2}_{{1}}&q^{1,2}_{{2}}&q^{1,3}_{{1}}&q^{1,3}_{{2}}
&q^{1,3}_{{3}}\\ \hdashline[2pt/2pt]
q^{2,1}_{{1}}&0&q^{2,2}_{
{1,1}}&0&q^{2,3}_{{1}}&q^{2,3}_{{2}}&0\\
0& q^{2,1}_{{1}}&0&q^{2,2}_{{1}}&0&q^{2,3}_{{1}}&q^{2,3}_{{2}}\\ \hdashline[2pt/2pt]
q^{3,1}_{{1}}&0&q^{3,2}_{{1}}&0&
q^{3,3}_{{1}}&q^{3,3}_{{2}}&0\\
0&q^{3,1}_{{1
}}&0&q^{3,2}_{{1}}&0&q^{3,3}_{{1}}&q^{3,3}_{{2}}
\\ \hdashline[2pt/2pt]
q^{4,1}_{{1}}&0&q^{4,2}_{{1}}&0&q^{4,3}
_{{1}}&q^{4,3}_{{2}}&0\\
0&q^{4,1}_{{1}}&0
&q^{4,2}_{{1}}&0&q^{4,3}_{{1}}&q^{4,3}_{{2}}
\\ \hdashline[2pt/2pt]
q^{5,1}_{{1}}&0&q^{5,2}_{{1}}&0&q^{5,3}
_{{1}}&q^{5,3}_{{2}}&0\\
0&q^{5,1}_{{1}}&0
&q^{5,2}_{{1}}&0&q^{5,3}_{{1}}&q^{5,3}_{{2}}
\end {array}
 \right]
\]
\noindent
For (\ref{nequ11_2}) to be solvable it is necessary to have
$q^{1,1}_{ 2} = 0$, $q^{1,2}_{ 2} = 0$, $q^{1,3}_{ 3} = 0$,
$q^{3,1}_{1} = 0$, $q^{3,2}_{1} = 0$, $q^{3,3}_{ 2} = 0$. With this selection of parameters the row highest order coefficient of $\hat{C} \tilde S(s) diag \{s,s,1\}$ is
\[
N_{\alpha}=
\left[ \begin {array}{ccc} q^{4,1}_{{1}}+q^{5,1}_{{1}}&q^{4,2}_{{1}}+q^{5,2}_{{1}}&q^{4,3}_{{2}}+q^{5,3}_{{2}}
\\ \noalign{\medskip}q^{2,1}_{{1}}+q^{5,1}_{{1}}&q^{2,2}_{{1
,1}}+q^{5,2}_{{1}}&q^{2,3}_{{2}}+q^{5,3}_{{2}}
\\ \noalign{\medskip}q^{5,1}_{{1}}&q^{5,2}_{{1}}&q^{5,3}_{{2}}\end {array} \right]
\]
\noindent
and has full row rank, while
\begin{equation*}
\left[\tilde D(s) \right]_{hc}=
\left[ \begin {array}{ccc}
q^{2,1}_{{1}}&q^{2,2}_{{1}}&q^{2,3}_{{2}}\\
q^{4,1}_{{1}}&q^{4,2}_{{1}}&q^{4,3}_{{2}}\\
q^{5,1}_{{1}}&q^{5,2}_{{1}}&q^{5,3}_{{2}}
\end {array} \right]
\end{equation*}
\noindent
and $Q_B$ also have full generic column rank and therefore the squared down system is decouplable. For further simplification we set
$q^{4,1}_{1} = 1$,  $q^{5,2}_{1} = 1$,  $q^{5,3}_{2} = 1$,
$q^{1,2}_{1} = 0$,  $q^{1,3}_{1} = 0$,  $q^{1,3}_{2} = 0$,
$q^{1,1}_{1} = 1$,  $q^{2,1}_{1} = 0$,
$q^{2,2}_{1} = 1$,
$q^{3,3}_{1} = 1$,
$q^{5,1}_{1} = 0$,
$q^{2,3}_{1} = 0$,  $q^{2,3}_{2} = 0$,
$q^{4,2}_{1} = 0$,  $q^{4,3}_{1}=  0$,
$q^{4,3}_{2} = 0$,
$q^{4,3}_{1} = 0$,
$q^{5,3}_{1} = 0$.
Then, by appropriately selecting $Q_A$ such that $[Q_A,Q_B]$ is invertible we get
\[
Q=
\left[Q_A,Q_B\right]=
\left[ \begin {array}{cc|ccccccc} 0&0&1&0&0&0&0&0&0
\\ 0&0&0&0&1&0&0&0&0\\ 0&0&0&0&0&1
&0&0&0\\ 0&0&0&0&0&0&1&0&0\\ 0&0&0
&0&0&0&0&1&0\\ 1&0&1&0&0&0&0&0&0
\\ 0&0&0&1&0&0&0&0&0\\ 0&1&0&0&1&0
&0&1&0\\ 0&0&0&0&0&1&0&0&1\end {array} \right]
\]
\noindent
The solution of (\ref{nequ13}) gives
\[\underline \mu^1(s)=
 \left[ \begin {array}{ccccccccc}
 s-t_1&-t_2&0&0&-t_2&t_1&-1&t_2&0
\end {array} \right], \ \underline \mu^2(s)=
 \left[ \begin {array}{ccccccccc}
 -t_3&-t_4&1&0&s-t_4&t_3&0&t_4&-1
\end {array} \right]
\]
\noindent
where $t_1$,$t_3$,$t_3$, and $t_4$ are free parameters.  Then
\[
F_0=
\left[ \begin {array}{ccrccccrc} t_1&t_2&0
&0&t_2&-1-t_1&1&-t_2&0
\\ 0&0&0&0&0&0&0&0&0\\ t_3&t_2&-2&0&t_4&-t_3&1&-
t_4&1\\ 0&0&0&0&0&0&0&0&0
\\ 0&0&0&0&0&0&0&0&0\end {array} \right],
G_0=
 \left[ \begin {array}{ccc} 0&0&0\\ 1&0&0
\\0&0&0\\ 0&1&0
\\ 0&0&1\end {array} \right]
\]
\noindent
Finally, after forming system (\ref{mpn2}), reordering the rows equations to get system $(A_r,B_r,C_r)$  as in Example \ref{example1} and applying similarity transformation $Q$   we get the squared down system to be decoupled
\[
\left[
\begin {array}{c|c}
sI-A_f&B_f\\ \hline
C_f&0
\end {array}
\right]
=
 \left[ \begin {array}{cc;{2pt/2pt}crcrcrc|rcc} s-t_1&-t_2&0&0&0&0&0&0&0&0&0&0\\
 -t_3&s-t_4&0&0&0&0&0&0&0&0&0&0\\ \hdashline[2pt/2pt]
 t_1&t_2&s&-1&0&0&0&0&0&0&0&0\\
 0&0&0&
s&0&0&0&0&0&0&1&0\\
0&0&0&0&s&-1&0&0&0&0&0&0
\\
0&0&0&0&1&s&0&0&0&1&0&0\\
0&0&0
&0&0&0&s&-1&0&0&0&0\\
t_3&t_4&0&0&0&0&0&s&-1&0&0&0\\
0&0&0&0&-1&1&0&0&s+1&-1
&0&1\\ \hline
0&1&0&1&2&1&0&1&1&0&0&0\\
0
&0&1&0&0&2&1&0&1&0&0&0\\
0&0&1&0&0&1&0&0&1&0&0&0
\end {array} \right]
\]
\noindent
Clearly, the input decoupling zeros are determined by the top left block of the above matrix (they are the roots of the polynomial $t(s)=s^2+(-t_1-t_4)s+t_4t_1-t_3t_2$
) and can be set by appropriate selection of the free parameters $t_1,t_2,t_3$ and $t_4$. The feedback pair
\[
 F_f=
 \left[ \begin {array}{ccrrrrrrr} 0&0&-2&0&1&-5&-3&-1&-2
\\ 0&0&1&-9&-20&-11&0&-10&-10\\ 0&0
&-1&-1&0&0&0&0&0\end {array} \right],
G_f=
 \left[ \begin {array}{ccr} 0&1&-1\\ 1&0&-1
\\ 0&0&1\end {array} \right]
\]
\noindent
results in the diagonal closed loop transfer function
\[
H_{cl}(s)= \mbox{diag} \Biggl\{  \left( s+10  \right)^{-1}, \left(s+3\right)^{-1}, \left(s+1 \right)^{-1} \Biggr\}
\]
\noindent
The decoupling pair in terms of the original system is
\[
F=
 \left[ \begin {array}{rrrrrcrrr} t_1&t_2&0
&0&t_2&-1-t_1&1&-t_2&0
\\ -2&1&-3&-3&-1&0&0&0&-2\\ t_3&t_4&-2&0&t_4&-t_3&1&-t_4&1\\
1&-20&-1&0&-10&0&-9&0&-
10\\ -1&0&0&0&0&0&-1&0&0\end {array} \right], \
G=
 \left[ \begin {array}{ccr} 0&0&0\\ 0&1&-1
\\0&0&0\\ 1&0&-1
\\ 0&0&1\end {array} \right]
\]
\end{exmpl}

\section{Conclusions}
Morgan's problem is solved by using an approach based on the
effect of singular input transformations on the input--state pencil of the original system. Such transformations split the set of resulting square  systems into a finite number of families characterised by common sets of column minimal indices of the input--state pencil or, equivalently, controllability indices and decouplability properties. The method proposed, is an exhaustive search for decouplable families according to the above classification. The decouplability is tested by using criteria from the  decoupling of square systems. The search procedure is described in detail and the algorithm for the problem solution is outlined. The problem of fixed poles, when a solution exists, is also considered and it is shown that when the sum of controllability indices is less than the system dimension the input decoupling zeros can be assigned.

\bibliographystyle{plain}

\end{document}